\documentclass[prl, twocolumn, superscriptaddress]{revtex4}
\usepackage[english]{babel}
\usepackage[dvips]{graphicx}
\usepackage{epsfig}
\usepackage{psfrag}
\usepackage{amsmath}
\usepackage{amssymb}
\usepackage{subfigure}
\begin{document}
\title{Electronic superlattices in corrugated graphene}
\author{A. Isacsson}
\email{andreas.isacsson@chalmers.se}
\affiliation{Department of Applied Physics, Chalmers University of
Technology, SE-412 96 G{\"o}teborg, Sweden}
\author{L. M. Jonsson}
\affiliation{Department of Applied Physics, Chalmers University of
Technology, SE-412 96 G{\"o}teborg, Sweden}
\author{J. M. Kinaret}
\affiliation{Department of Applied Physics, Chalmers University of
Technology, SE-412 96 G{\"o}teborg, Sweden}
\author{M. Jonson}
\affiliation{Department of Physics, G\"{o}teborg University, SE-412
96 G\"{o}teborg, Sweden} \affiliation{School of Engineering and
Physical Sciences, Heriot-Watt University, Edinburgh EH14 4AS,
Scotland, UK}
\begin{abstract}
  We theoretically investigate electron transport through corrugated
  graphene ribbons and show how the ribbon curvature leads to an
  electronic superlattice with a period set by the corrugation wave
  length. Transport through the ribbon depends sensitively on the
  superlattice band structure which, in turn, strongly depends on the
  geometry of the deformed sheet.  In particular, we find that for
  ribbon widths where the transverse level separation is comparable to
  the the band edge energy, a strong current switching occurs as
  function of an applied backgate voltage. Thus, artificially
  corrugated graphene sheets or ribbons can be used for the study of
  Dirac fermions in periodic potentials. Furthermore, this provides an
  additional design paradigm for graphene-based electronics.
\end{abstract}

\maketitle

A single layer of graphite, known as graphene,
was for the first time studied experimentally in 2004
\cite{NovoselovGeim04}. This started a massive interest in graphene,
mainly
because it is a 2D gapless semiconductor with massless
``relativistic'' quasiparticles
\cite{NovoselovGeim05,ZhangTan05,NomuraMacDonald07,TworzydloTrauzettel06,GeimNovoselov07,KatsnelsonNovolseov07,PargaCalleja07}.
An unusual integer quantum Hall effect \cite{GusyninSharapov05} and
a predicted minimal conductivity $\sigma =4e^2/(h \pi)$ are among
the manifestations of 
the linear energy dispersion
in
the vicinity of the Fermi energy.  In addition to
being a tool for studying fundamental physics, graphene is also of
interest for device applications (\emph{e.g.}  transistors, lenses
and NEMS resonators \cite{CheianovFalko06, CheianovFalko07,
XuZheng07,BunchvanderZande07}).

Proposed graphene-based devices typically rely on external
electrostatic gates for controlling the electronic transport.  In this
Letter, however, we show that the effective potential induced in a
graphene ribbon placed on a corrugated substrate, can strongly alter
the transport properties of graphene. This effective potential is
determined by the local curvature~\cite{CastroNetoKim07} of the ribbon
and introduces an additional design degree of freedom of interest for
both fundamental studies of graphene and graphene-based electronic
devices. For illustration we focus here on periodically modulated
surfaces --- superlattices --- and show how the corresponding band
structure can be readily probed by conductance measurements.

\begin{figure}[t]
\epsfig{file=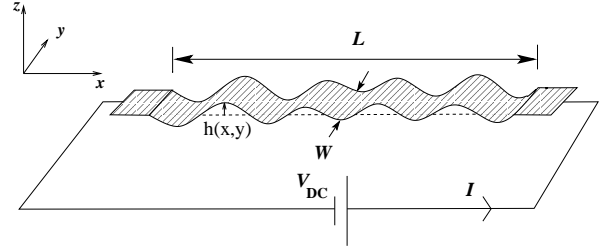,width=0.9\linewidth}\vspace*{-0.2cm}
\caption{Schematic illustration of the graphene ribbon system
investigated in this paper. The
DC-biased ribbon of width $W$ rests on a corrugated substrate
surface of length $L$ and height profile $h(x,y)$. }\vspace*{-0.2cm}
\label{fig:modelsystem}
\end{figure}
We consider a single graphene ribbon placed on a
corrugated surface and biased by a small applied DC voltage as shown
in Fig.~1. In the tight binding description the ribbon is described by
\begin{equation}
\hat H_{{\rm TB}}=\sum_{\left<i,j\right>}t_{ij}a^\dagger_ib_j+\sum_{\left<\left<i,j\right>\right>}t_{ij}^\prime\left(a^\dagger_i
a_j+b^\dagger_ib_j\right)+{\rm h.c.}
\label{eq:tbhamiltonian}
\end{equation} 
Here $\left<i,j\right>$ denote nearest neighbors ($a$ and $b$ atoms)
and $\left<\left<i,j\right>\right>$ next nearest neighbors.  When the
ribbon is deformed, the matrix elements $t$ and $t'$ change.  Thus, if
$t_{ij}=V_{pp\pi}$ describes hopping between atoms $i$ and $j$ on a
flat graphene sheet, we find \cite{FootNote} that for a deformed
sheet~\cite{CastroNetoKim07,Harrison04} $t_{ij}^{(\prime)}$ is replaced by
$$\tilde{t}_{ij}^{(\prime)}=\frac{u_{ij}^2}{d_{ij}^4}\left[(V_{pp\sigma}
  \!  -V_{pp\pi} \! )({\bf n}_i \! \cdot \! {\bf d}_{ij})({\bf n}_j \!
  \cdot \!{\bf d}_{ij})\!+ \!V_{pp\pi}d_{ij}^2{\bf n}_i \! \cdot \!
	{\bf n}_j\right].$$ Here ${\bf u}_{ij}$ is the vector
connecting atoms $i$ and $j$ in the undeformed lattice while ${\bf
  d}_{ij}$ is the corresponding vector after deformation.  The surface
normals are denoted by ${\bf n}_{i(j)}$.  While a general deformation
involves both bending and stretching, we will here restrict our
attention to pure bending deformations in one direction, {\em i.e.}
$z=h(x)$ (see Fig.~1). We write the new matrix elements
$\tilde{t}_{ij}^{(\prime)}=t_{ij}^{(\prime)}+\delta_{ij}^{(\prime)}$,
and to second order in $\partial_x^2 h$ we find
$$\delta_{ij}^{(\prime)}=-\frac{(\partial^2_xh)^2({\bf
u}_{ij}\cdot\hat{x})^4}{2u_{ij}^2}\!\left[\!\left(\!\frac{u_{ij}^2}{({\bf
u}_{ij}\cdot\hat{x})^2}\!-\!\frac{2}{3}\!\right)\!
V_{pp\pi}+\frac{1}{2}V_{pp\sigma}\right].$$ We note that this
expression is similar but not identical to the one in
Ref.~\cite{CastroNetoKim07}. Inserting the new matrix elements in
Eq. (\ref{eq:tbhamiltonian}) and expanding around the Fermi points
${\bf K}$ and ${\bf K}'$, results in an effective Hamiltonian with two
new terms ${\bf A}_{\rm eff}$ and $\Phi_{{\rm eff}}$ which are of
order $a_0^2(\partial_x^2 h)^2$ ($a_0=1.42$ {\AA} is the lattice
constant). For electrons near the ${\bf K}$ point the Hamiltonian is
\begin{equation}
\hat H_{{\rm eff}}=\hbar v_F\hat{\sigma}\cdot\left[-i\nabla+{\bf
A}_{\rm eff}({\bf x)}\right]+eV_G+\Phi_{{\rm eff}}({\bf x}).
\label{eq:dirachamiltonian}
\end{equation}
Here $\hat \sigma = \hat x \sigma_x + \hat y \sigma_y$ and
$\sigma_{x,(y)}$ are Pauli matrices, $v_F=10^6$ m/s is the Fermi
velocity, and $eV_G$ can be produced by the action of a back-gate.

Unless time reversal symmetry is broken, {\em e.g.} by a magnetic
field, the effective vector potential ${\bf A}_{\rm eff}({\bf x})$ at
the ${\bf K}$-point, and its time reversed counterpart at the ${\bf
K}'$-point, only contribute to second order in $a_0^2(\partial_x^2
h)^2$ and will be ignored in what follows.  The second new term
corresponds to an effective potential, which for graphene bent along
the ``armchair'' direction is
$$\Phi_{\rm eff}(x)=\frac{27}{4}a_0^2(\partial_x^2
h)^2\left(\frac{3}{8}V_{pp\sigma}^{(aa)}-\frac{1}{6}V_{pp\pi}^{(aa)}\right).$$
In addition, because of the vanishing of the first order matrix
element of ${\bf A}_{\rm eff}({\bf x})$, a third new term
corresponding to a local variation in Fermi-velocity should be
considered. However, for long wavelengths, $ka_0 \ll 1$, this term can
be shown to be much smaller than the effective potential $\Phi_{{\rm
eff}}$. 

The form of the effective potential is simplified if we take the shape of
the ribbon to be $h(x,y) = A\sin(n \pi x/L)$. This approximation
captures the qualitative behavior of a general periodic potential and
results in the expression
\begin{equation}
\Phi_{\rm eff}(x)=E_0(A/a_0)^2(k_{\rm s}a_0)^4(1-\cos k_{\rm s}x)\,,
\label{eq:Phi}
\end{equation}
where $k_{\rm s}=2n\pi/L$ and $E_0\approx 0.22$ eV. Note that
$\Phi_{\rm eff}(x)$ is positive definite (repulsive), and its strength
varies rapidly with $k_{\rm s}$. Because of this, large amplitudes
$A/a_0 \gg 1$ are neccesary. Hence, it is important that
Eq. (\ref{eq:Phi}) is transformed to a coordinate system $s=s(x)$ that
follows the graphene ribbon.  The relation between $s(x)$ and $x$ is
$$s(x) = \frac{L}{\pi n} \sqrt{1+ \tilde A^2} E \bigg(\frac{n \pi
x}{L},\sqrt{\frac{\tilde A^2}{1 + \tilde A^2}}\bigg),$$ where $E$ is
the elliptic integral of the second kind and $\tilde A = n \pi A/L$.
The total length of the graphene sheet is then $s(L)$. For
simplicity we will from here on write $x$ rather than $s(x)$ for the
coordinate along the sheet.

For narrow graphene nano-ribbons the choice of transverse electronic
boundary conditions is of great importance. They depend on the
configuration of carbon atoms along the edge \cite{BreyFertig06} as
well as the ribbon width.  In this Letter, we use boundary conditions
for a metallic armchair edge \cite{FootNote}.  The wave vector
quantization in the transverse direction ($y$) gives $k_n =
n{\pi}/{W}$.  In the absence of fields the wave functions satisfy the
Dirac equation
\begin{eqnarray}
-i\hbar v_{\rm F}\left[\begin{tabular}{lr}
    $\nabla\cdot\hat{\sigma}$ & 0 \\
    0 & $\nabla\cdot\hat{\sigma}^*$
\end{tabular}\right]e^{ikx}\phi_n^\pm(k,y)=\epsilon_n^\pm
e^{ikx}\phi_n^\pm(k,y) \nonumber
\end{eqnarray}
with $\epsilon^\pm_n=\pm \hbar v_F\sqrt{k^2+k_n^2}$.  For $n>0$ the
eigenspinors are 
\begin{eqnarray}
\phi_n^{\pm}(k,y)=e^{ik_ny}\left(\begin{tabular}{c}
$1$\\
$\pm e^{i\phi_n(k)}$\\
$\pm e^{i\phi_n(k)}$\\
$1$
\end{tabular}\right)+
e^{-ik_ny}\left(\begin{tabular}{c}
$\pm e^{i\phi_n(k)}$\\
$1$\\
$1$\\
$\pm e^{i\phi_n(k)}$
\end{tabular}\right),\nonumber
\end{eqnarray} 
which are two-fold degenerate and for $n=0$ they are  
$$\phi_0^{\pm}(k,y)=[1, \pm {\rm sgn}(k), 1, \pm{\rm sgn}(k)]^T,$$
and non-degenerate. The factors
$\exp(i\phi_n(k))=(k+ik_n)/\sqrt{k^2+k_n^2}$ correspond to the
incidence angle of the electrons.

In the presence of a scalar potential that only depends on $x$ there
can be no band-mixing although positive and negative energy
solutions belonging to the same band may mix. Thus we look for
solutions of the form
\begin{eqnarray}
\psi_n(x,y)\!&=&\!\int dk\,
\gamma_n^+(k)\left[\phi_n^+(k,y)\!+ \!\phi_n^-(k,y)\right]e^{ikx}\nonumber\\
\!&+&\! \int
dk\,\gamma_n^-(k)\left[\phi_n^+(k,y)\! -\! \phi_n^-(k,y)\right]e^{i[kx-\phi_n(k)]}\nonumber,
\end{eqnarray}
which, together with Eq. \ref{eq:dirachamiltonian}, leads to the equation
\begin{equation}
\left[-i\hbar v_F(\partial_x\sigma_x
+ik_n\sigma_y)+V(x)\right]\bar{\gamma}_n(x)=\epsilon
\bar{\gamma}_n(x) \label{eq:1D}
\end{equation}
for the two-component spinor
$\bar{\gamma}_n(x)=[\gamma_n^+(x),\gamma_n^-(x)]$ with $V(x) = eV_{\rm
G} + \Phi_{{\rm eff}}$.

\begin{figure}
\includegraphics[width=0.9\linewidth,clip]{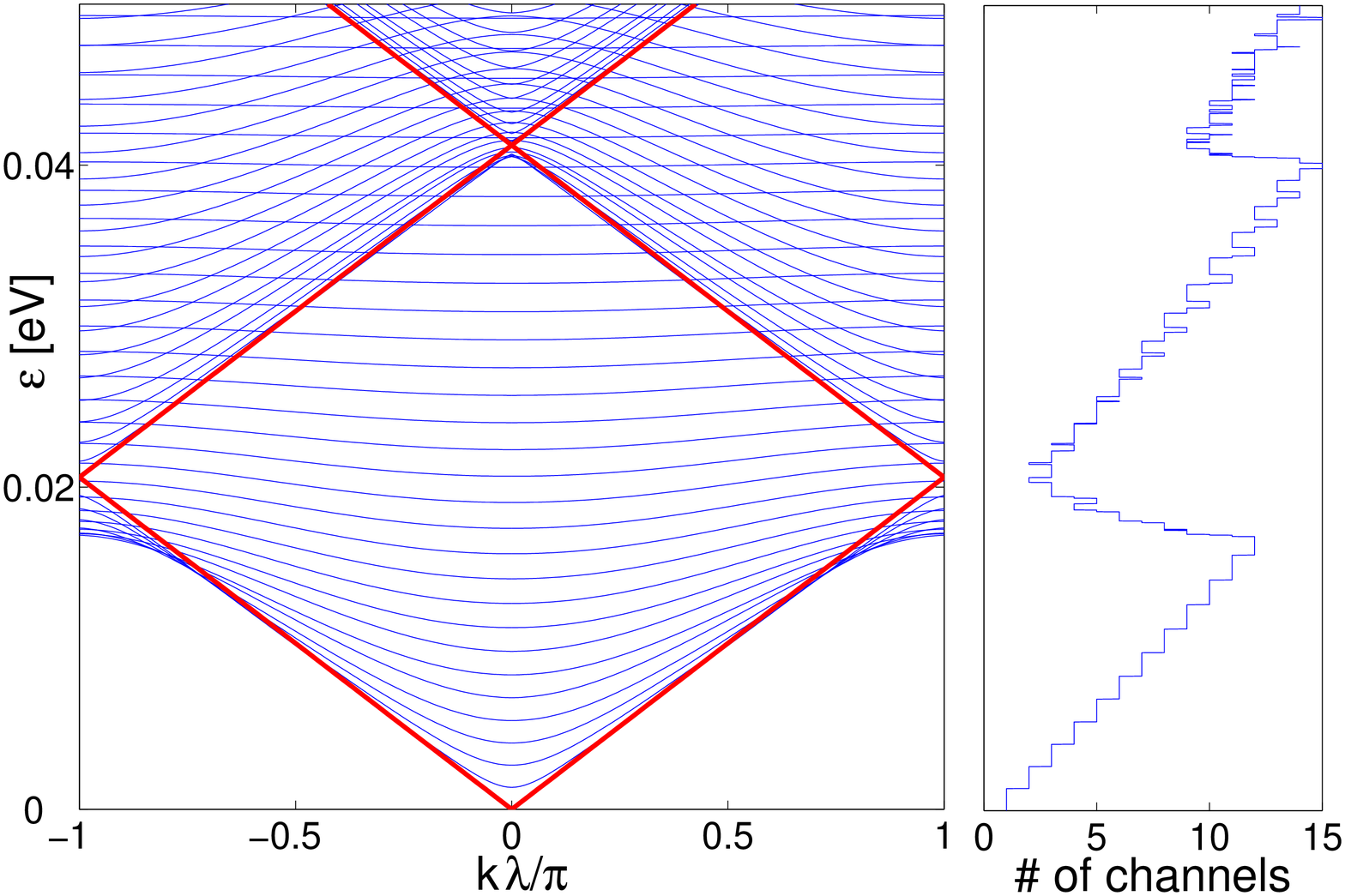}\vspace*{-0.2cm}
\caption{ (Color online) Band structure for an infinitely long and 1
$\mu$m wide metallic graphene sheet with armchair edges subjected to
a periodic potential $V_0 \sin (2 \pi x/ \lambda)$ ($V_0 \sim 2.5$
meV) (left panel) and the corresponding number of bands crossing
each energy (right panel). The linear band calculated for $k_n =0$
(bold line) is not distorted by the periodic potential, which is a
manifestation of the Klein paradox. A pseudogap is visible for $E
\sim 0.02$ eV and another at $E \sim 0.04$ eV roughly corresponding
to the free electron energies at $k = \pi \lambda$ and $k=2\pi
\lambda$. The band structure is in this case symmetric for positive
and negative energies (not shown).}\vspace*{-0.2cm}
\label{fig:bands}
\end{figure}

Before analyzing the curvature effects we begin with a general
discussion of the band structure for periodic potentials
in the Dirac equation (see also Refs.
\cite{RoyMendez94,SamsonovPecheritsin03}). For this we consider an
infinite graphene strip of width $W$ subject to the periodic potential
$V(x) = V_0 \sin(2 \pi x/L)$. 
In this case Eq. (\ref{eq:1D}) can be be solved numerically and the
resulting band structure in the reduced zone scheme is shown in the
left panel of Fig. \ref{fig:bands}.  The right panel shows the number
of conducting channels as a function of energy and one sees that the
periodic potential produces pseudo gaps whenever this number has a
minimum.  It is interesting to note that the gap sizes increase with
band index $n$. This is consistent with our knowledge of Klein
tunnelling since a potential barrier has no effect on massless
relativistic particles while for finite mass there is an effect which
increases with mass \cite{CalogeracosDombey99}.  Here we consider
one-dimensional motion along the graphene ribbon in bands
corresponding to quantized transverse momenta $k_n$. In the effective
equation for the longitudinal motion, Eq. (\ref{eq:1D}), these
transverse momenta produce a mass term that is zero for $n=0$ and
finite and increasing with $n$ for $n>0$.
\begin{figure}[t]
\centering
\includegraphics[width=0.85\linewidth,height=0.6\linewidth,clip]{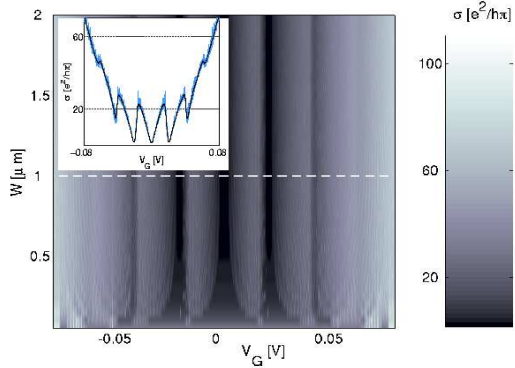}\vspace*{-0.2cm}
\caption{(Color online) Conductivity ($\sigma=GL/W$) for a graphene
sheet of length $L = 1$ $\mu$m with a periodic potential $V_0 \sin (2
\pi x /\lambda)$ ($V_0 \sim 2.5$ meV, $\lambda = 100$ nm) as a
function of gate voltage and sheet width $W$. The pseudo gaps at $V_G
= 0.02$ eV and $V_g = 0.04$ eV shown in Fig. \ref{fig:bands}
correspond to local minima in the conductivity. The inset shows the
conductivity along the white dashed line ($W=1$ $\mu$m) corresponding
to the band diagram in Fig.~\ref{fig:bands}.  The thick line
represents an average over nearby gate voltages, which removes the
Fabry-Perot interferences (see text). The non-averaged data is shown
as the thin (blue) line.}\vspace*{-0.2cm}
\label{fig:conductance}
\end{figure}
\begin{figure}[t]
\includegraphics[width=\linewidth]{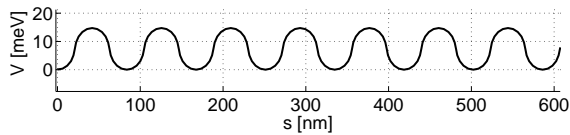}\vspace*{-0.2cm}
\caption{Effective potential along a graphene sheet of a shape
defined by $h(x) = A \sin( 2 \pi x /\lambda)$; $A= 20$ nm, $\lambda
= 20$ nm.}\vspace*{-0.2cm} \label{fig:potential}
\end{figure}

The band structure in Fig. \ref{fig:bands} can be probed by transport
measurements. To demonstrate this we adopt a Landauer approach
together with a transfer matrix method and calculate the transmission
probabilities assuming coherent and ballistic transport.  The
conductance is then found from $G = (4e^2/h) \sum_n t_n$ where $t_n$
is the transmission probability for channel $n$ and the sum over $n$
runs over all channels. The transmission probabilities $t_n$ are
obtained from the transfer matrices $T_n = \prod_{m=1}^N
T_n^{(m)}$ ~\cite{Datta95}.  Here the interval $0<x<s(L)$ is divided
into $N$ steps each having a constant potential $V_m=V(x_m)$, $x_m =
m s(L)/N$. The transfer matrices between slices are found from the
requirement that the wave functions be continuous everywhere (current
conservation). The reservoirs on the left and right sides are taken to
be infinitely wide graphene strips. As will be seen, this gives rise
to a Fabry-Perot like interference pattern in the conductance, due to
reflections at the reservoir-ribbon interfaces, as the back gate
voltage is varied.  These fluctuations are expected to smear out at
finite temperatures and in the presence of impurities.  As pointed out
in Ref.  \cite{Katsnelson07} it is in general not allowed (as in the
case of a 2DEG) to introduce a general adiabatic widening of the strip
to remove these fluctuations. Such a widening will introduce a
changing structure of the transverse boundary conditions and a
detailed description of the edge geometry is necessary.

The conductivity of a finite system of length $L=1\,\mu$m is shown as
a function of gate voltage and strip width in
Fig.~\ref{fig:conductance}.
For widths $W>L\gg \lambda$ the conductivity does not change
appreciably as the the ribbon becomes wider whereas for narrow strips
strong alteration of the conductivity occurs. The inset shows the
conductivity for parameters corresponding to Fig.~\ref{fig:bands}. The
conductivity minima agree with the predicted pseudo gaps from the
infinite structure in Fig. \ref{fig:bands}.

Now we consider a finite length graphene ribbon placed on a corrugated
substrate.  We will specifically consider narrow strips where the
transverse energy level spacing $\Delta \epsilon_n\sim \hbar v_F/W$ is
of the same order as the band edge energy, i.e. $\lambda\sim W$.  For
illustration we chose a ribbon placed on a sinusoidally shaped
substrate, $h(x)=A\sin(2\pi x/\lambda)$ with $A=20$ nm and $\lambda=
20$ nm. This leads to an effective potential (see
Fig.~\ref{fig:potential}) of the order of 10 meV with an effective
wavelength of the order of 80 nm (consistent with the assumption $k
a_0 \ll 1$). Choosing a width of $W=100$~nm thus puts us in the
desired regime. Figure \ref{fig:ribbon} shows the conductance of a
graphene ribbon with $L=1$ $\mu$m as a function of the back-gate
voltage (thick blue solid curve) calculated using the transfer matrix
method described above. The conductance of a flat ribbon of equivalent
length $s(L)$ is also shown for comparison (red dashed curve). Both
curves have been averaged over nearby points to remove spurious
interferences arising from the abrupt boundary
conditions. Non-averaged data for the corrugated sheet is shown as the
thin (blue) line.

A comparison of the conductances of flat and corrugated graphene
sheets shown in Fig.~\ref{fig:ribbon} reveals two distinct features.
Firstly, there is an asymmetry in the conductance of a corrugated
sheet with respect to positive and negative back-gate voltages because
the average effective potential $\bar \Phi_{{\rm eff}}$ is strictly
positive. This results in a total shift $\Delta V_G = \bar \Phi_{{\rm
eff}}$ as well as in changes in the detailed structure.  Secondly, the
effect of corrugation is clearly seen to strongly alter the
conductance due to conductance channels being switched on and
off. This can be traced to the band structure for an infinite system
with corresponding parameters shown in the right panel of
Fig. \ref{fig:ribbon} (\emph{N.B.} the unaffected $n=0$ band has been
omitted).

\begin{widetext}
\vspace*{-1.0cm}
\mbox{}
\begin{figure}[t]
\includegraphics[width=12cm, height=4.5cm,clip]{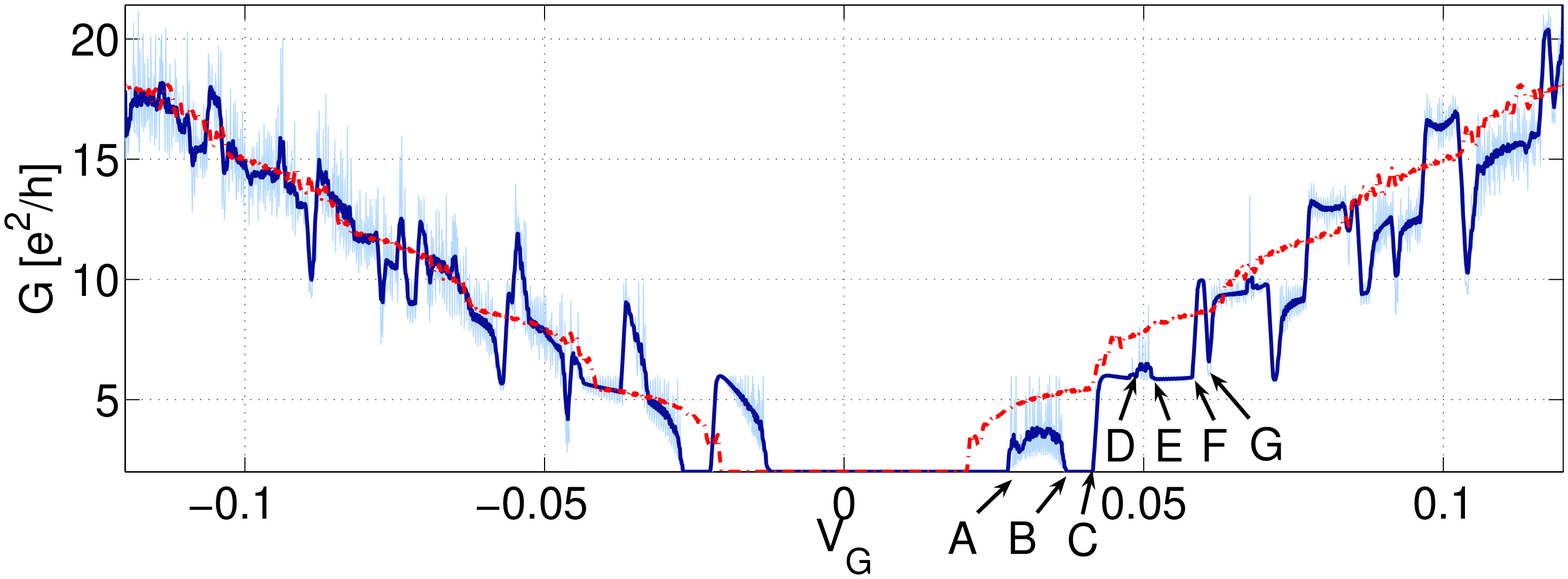}
\hspace{-1.0cm} \raisebox{0.3cm}
{\includegraphics[width=4cm,height=3.9cm,clip]{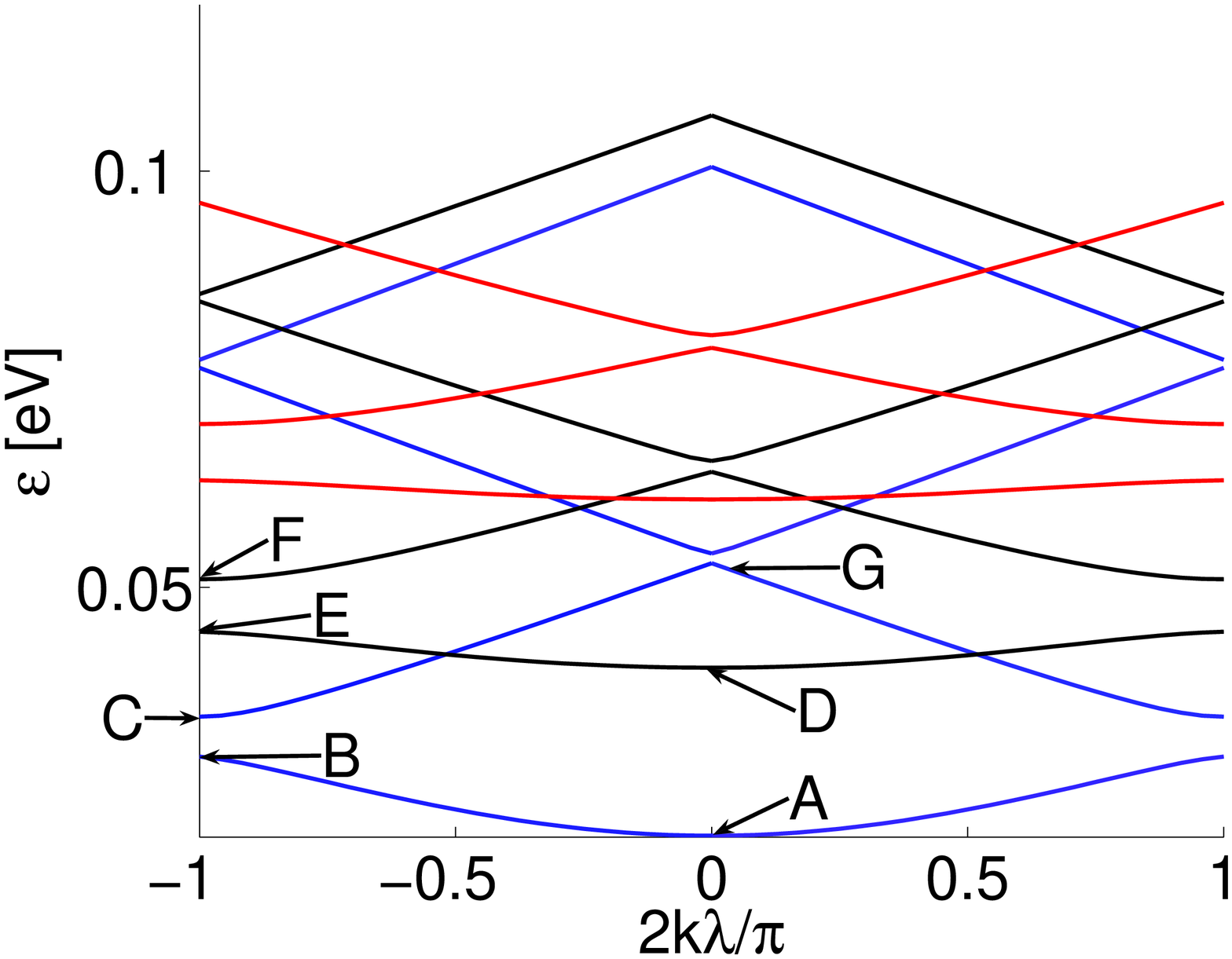}}
\vspace*{-0.4cm}
\caption{(Color online) Conductance of a graphene ribbon as a function
of gate voltage (left panel). The conductance of a ribbon placed on a
corrugated surface (thick blue solid curve) is asymmetric with respect
to the sign of the gate voltage in contrast to when the surface is
flat (red dashed curve). The origin of this asymmetry is that the
curvature-induced potential is not symmetric. Focusing on positive
gate bias we can identify points (A-G) in the conductance curve,
corresponding to switching on and off conductance in specific
channels.  The corresponding band edges for an infinite ribbon are
shown in the right panel. For negative $V_{\rm G}$ similar features
appear in the conductance (not labelled).}\vspace*{-0.4cm}
\label{fig:ribbon}
\end{figure}
\end{widetext}
Finally, we have also considered short graphene ribbons where $L\sim
W\sim\lambda$.  Figure~\ref{fig:short} shows the conductance of a
ribbon with $s(L)\approx 160$ nm and therefore only two potential
maxima (double barrier). Again, comparing with a flat system of equal
length we find that the overall features in the conductance change in
the same qualitative ways as described above. In this case, however,
one should be careful not to confuse structure in the conductance due
to the Fabry-Perot like interferences on the one hand with structure
due to resonant tunneling on the other. In this case both effects are
of similar importance.
\begin{figure}[]
\vspace*{-0.4cm}
\includegraphics[width=\linewidth]{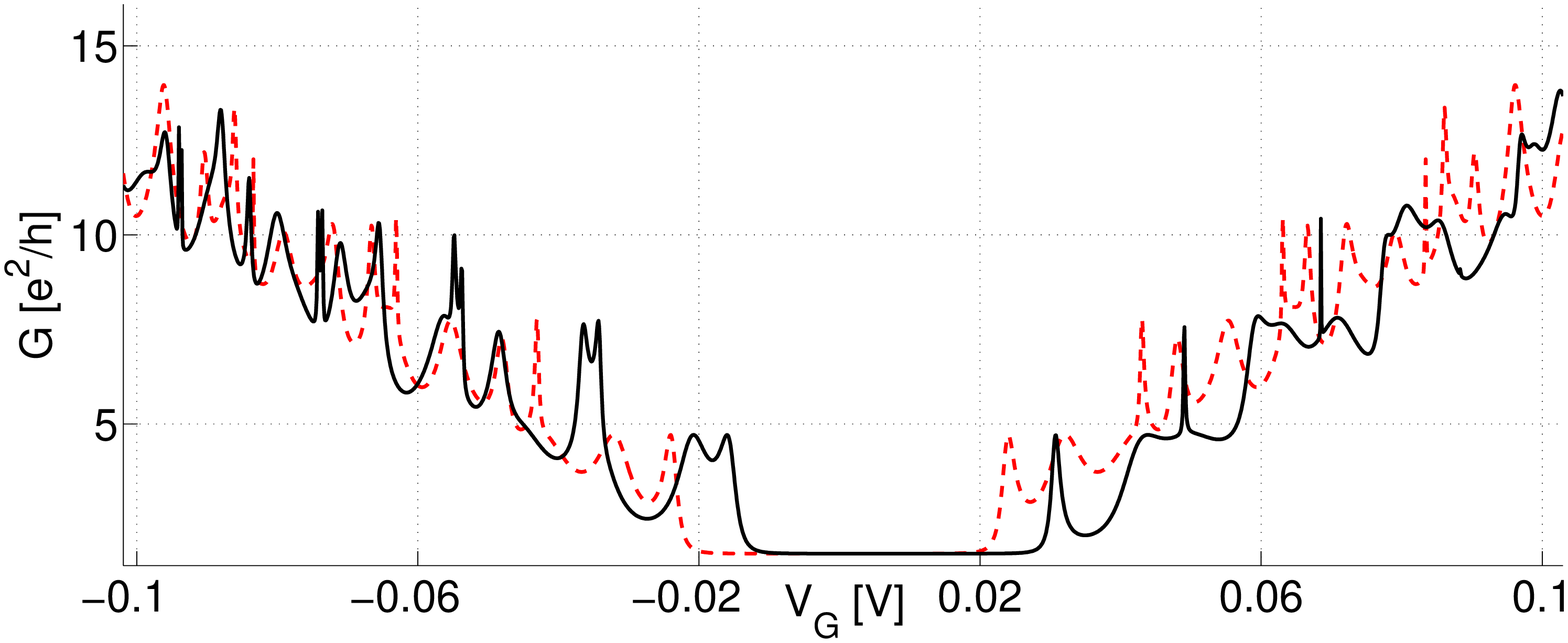}\vspace*{-0.2cm}
\caption{(Color online) Conductance of a short metallic ribbon with
(solid, black) and without (dashed, red)
corrugation.}\vspace*{-0.35cm} \label{fig:short}
\end{figure}

In conclusion, we have shown that placing graphene on an artificially
corrugated surface produces an effective local potential for the
graphene electrons. This potential, which is related to the local
curvature, can be tailored to significantly alter the transport
properties of graphene. Specifically we have considered the effect on
the electrical conductance of periodic potentials and showed how the
band structure manifest itself in graphene nano-ribbons. Such a
relation between transport properties and geometrical configurations
may add to the number of design degrees of freedom available for
constructing graphene based electronic devices. It may also provide an
alternative transduction mechanism in graphene based NEMS.

We are grateful for stimulating discussions with Leonid Gorelik.
Funding was provided by the Swedish Foundation for
Strategic Research (SSF).
\bibliography{}

\begin{thebibliography}{20}
\expandafter\ifx\csname natexlab\endcsname\relax\def\natexlab#1{#1}\fi
\expandafter\ifx\csname bibnamefont\endcsname\relax
  \def\bibnamefont#1{#1}\fi
\expandafter\ifx\csname bibfnamefont\endcsname\relax
  \def\bibfnamefont#1{#1}\fi
\expandafter\ifx\csname citenamefont\endcsname\relax
  \def\citenamefont#1{#1}\fi
\expandafter\ifx\csname url\endcsname\relax
  \def\url#1{\texttt{#1}}\fi
\expandafter\ifx\csname urlprefix\endcsname\relax\def\urlprefix{URL }\fi
\providecommand{\bibinfo}[2]{#2}
\providecommand{\eprint}[2][]{\url{#2}}

\bibitem[{\citenamefont{Novoselov et~al.}(2004)\citenamefont{Novoselov, Geim,
  Morozov, Jiang, Zhang, Dubonos, Grigorieva, and Firsov}}]{NovoselovGeim04}
\bibinfo{author}{\bibfnamefont{K.~S.} \bibnamefont{Novoselov} et~al. },
  \bibinfo{journal}{Science}
  \textbf{\bibinfo{volume}{306}}, \bibinfo{pages}{666} (\bibinfo{year}{2004}).

\bibitem[{\citenamefont{Geim and Novoselov}(2007)}]{GeimNovoselov07}
\bibinfo{author}{\bibfnamefont{A.~K.} \bibnamefont{Geim}} \bibnamefont{and}
  \bibinfo{author}{\bibfnamefont{K.~S.} \bibnamefont{Novoselov}},
  \bibinfo{journal}{Nat. Mater.} \textbf{\bibinfo{volume}{6}},
  \bibinfo{pages}{183} (\bibinfo{year}{2007}).

\bibitem[{\citenamefont{Katsnelson and
  Novoselov}(2007)}]{KatsnelsonNovolseov07}
\bibinfo{author}{\bibfnamefont{M.}~\bibnamefont{Katsnelson}} \bibnamefont{and}
  \bibinfo{author}{\bibfnamefont{K.}~\bibnamefont{Novoselov}},
  \bibinfo{journal}{Solid State Commun.} \textbf{\bibinfo{volume}{143}},
  \bibinfo{pages}{3} (\bibinfo{year}{2007}).

\bibitem[{\citenamefont{Nomura and MacDonald}(2007)}]{NomuraMacDonald07}
\bibinfo{author}{\bibfnamefont{K.}~\bibnamefont{Nomura}} \bibnamefont{and}
  \bibinfo{author}{\bibfnamefont{A.~H.} \bibnamefont{MacDonald}},
  \bibinfo{journal}{Phys. Rev. Lett.} \textbf{\bibinfo{volume}{98}},
  \bibinfo{eid}{076602} 
  (\bibinfo{year}{2007}).

\bibitem[{\citenamefont{Novoselov et~al.}(2005)\citenamefont{Novoselov, Geim,
  Morozov, Jiang, Katsnelson, Grigorieva, Dubonos, and
  Firsov}}]{NovoselovGeim05}
\bibinfo{author}{\bibfnamefont{K.~S.} \bibnamefont{Novoselov} et al.},
  \bibinfo{journal}{Nature}
  \textbf{\bibinfo{volume}{438}}, \bibinfo{pages}{197} (\bibinfo{year}{2005}).

\bibitem[{\citenamefont{Tworzydlo et~al.}(2006)\citenamefont{Tworzydlo,
  Trauzettel, Titov, Rycerz, and Beenakker}}]{TworzydloTrauzettel06}
\bibinfo{author}{\bibfnamefont{J.}~\bibnamefont{Tworzydlo}},
  \bibinfo{author}{\bibfnamefont{B.}~\bibnamefont{Trauzettel}},
  \bibinfo{author}{\bibfnamefont{M.}~\bibnamefont{Titov}},
  \bibinfo{author}{\bibfnamefont{A.}~\bibnamefont{Rycerz}}, \bibnamefont{and}
  \bibinfo{author}{\bibfnamefont{C.~W.~J.} \bibnamefont{Beenakker}},
  \bibinfo{journal}{Phys. Rev. Lett.} \textbf{\bibinfo{volume}{96}},
  \bibinfo{eid}{246802} 
  (\bibinfo{year}{2006}).

\bibitem[{\citenamefont{Zhang et~al.}(2005)\citenamefont{Zhang, Tan, Stormer,
  and Kim}}]{ZhangTan05}
\bibinfo{author}{\bibfnamefont{Y.}~\bibnamefont{Zhang}},
  \bibinfo{author}{\bibfnamefont{Y.-W.} \bibnamefont{Tan}},
  \bibinfo{author}{\bibfnamefont{H.~L.} \bibnamefont{Stormer}},
  \bibnamefont{and} \bibinfo{author}{\bibfnamefont{P.}~\bibnamefont{Kim}},
  \bibinfo{journal}{Nature} \textbf{\bibinfo{volume}{438}},
  \bibinfo{pages}{202} (\bibinfo{year}{2005}).

\bibitem[{\citenamefont{de~Parga et~al.}(2007)\citenamefont{de~Parga, Calleja,
  Borca, Jr., Hinarejos, Guinea, and Miranda}}]{PargaCalleja07}
\bibinfo{author}{\bibfnamefont{A.~L.} \bibnamefont{Vasquez~de~Parga} et al.},
  \bibinfo{journal}{arXiv:cond-mat/} \bibinfo{pages}{0709.0360},
  (\bibinfo{year}{2007}).

\bibitem[{\citenamefont{Gusynin and Sharapov}(2005)}]{GusyninSharapov05}
\bibinfo{author}{\bibfnamefont{V.~P.} \bibnamefont{Gusynin}} \bibnamefont{and}
  \bibinfo{author}{\bibfnamefont{S.~G.} \bibnamefont{Sharapov}},
  \bibinfo{journal}{Phys. Rev. Lett.} \textbf{\bibinfo{volume}{95}},
  \bibinfo{eid}{146801} 
  (\bibinfo{year}{2005}).

\bibitem[{\citenamefont{Bunch et~al.}(2007)\citenamefont{Bunch, van~der Zande,
  Verbridge, Frank, Tanenbaum, Parpia, Craighead, and
  McEuen}}]{BunchvanderZande07}
\bibinfo{author}{\bibfnamefont{J.~S.} \bibnamefont{Bunch} et al.},
  \bibinfo{journal}{Science}
  \textbf{\bibinfo{volume}{315}}, \bibinfo{pages}{490} (\bibinfo{year}{2007}).

\bibitem[{\citenamefont{Xu et~al.}(2007)\citenamefont{Xu, Zheng, and
  Chen}}]{XuZheng07}
\bibinfo{author}{\bibfnamefont{Z.}~\bibnamefont{Xu}},
  \bibinfo{author}{\bibfnamefont{Q.-S.} \bibnamefont{Zheng}}, \bibnamefont{and}
  \bibinfo{author}{\bibfnamefont{G.}~\bibnamefont{Chen}},
  \bibinfo{journal}{Appl. Phys. Lett.} \textbf{\bibinfo{volume}{90}},
  \bibinfo{eid}{223115} 
  (\bibinfo{year}{2007}).

\bibitem[{\citenamefont{Cheianov and Fal'ko}(2006)}]{CheianovFalko06}
\bibinfo{author}{\bibfnamefont{V.~V.} \bibnamefont{Cheianov}} \bibnamefont{and}
  \bibinfo{author}{\bibfnamefont{V.~I.} \bibnamefont{Fal'ko}},
  \bibinfo{journal}{Phys. Rev. B}
  \textbf{\bibinfo{volume}{74}}, \bibinfo{eid}{041403}
  (\bibinfo{year}{2006}).

\bibitem[{\citenamefont{Cheianov et~al.}(2007)\citenamefont{Cheianov, Fal'ko,
  and Altshuler}}]{CheianovFalko07}
\bibinfo{author}{\bibfnamefont{V.~V.} \bibnamefont{Cheianov}},
  \bibinfo{author}{\bibfnamefont{V.}~\bibnamefont{Fal'ko}}, \bibnamefont{and}
  \bibinfo{author}{\bibfnamefont{B.~L.} \bibnamefont{Altshuler}},
  \bibinfo{journal}{Science} \textbf{\bibinfo{volume}{315}},
  \bibinfo{pages}{1252} (\bibinfo{year}{2007}).

\bibitem[{\citenamefont{Neto and Kim}(2007)}]{CastroNetoKim07}
\bibinfo{author}{\bibfnamefont{A.~H.} \bibnamefont{Castro-Neto}} \bibnamefont{and}
  \bibinfo{author}{\bibfnamefont{E.-A.} \bibnamefont{Kim}}
  \bibinfo{journal}{arXiv:cond-mat/} \bibinfo{pages}{0702562},(\bibinfo{year}{2007}).

\bibitem[{\citenamefont{Fotnot}(2007)}]{FootNote}
\bibinfo{author}{More detailed derivations and results for other boundary conditions will be given elsewhere}.

\bibitem[{\citenamefont{Harrison}(2004)}]{Harrison04}
\bibinfo{author}{\bibfnamefont{W.~A.} \bibnamefont{Harrison}},
  \emph{\bibinfo{title}{Elementary Electronic Structure}}
  (\bibinfo{publisher}{World Scientific}, \bibinfo{address}{Singapore},
  \bibinfo{year}{2004}), \bibinfo{edition}{revised} ed.

\bibitem[{\citenamefont{Brey and Fertig}(2006)}]{BreyFertig06}
\bibinfo{author}{\bibfnamefont{L.}~\bibnamefont{Brey}} \bibnamefont{and}
  \bibinfo{author}{\bibfnamefont{H.~A.} \bibnamefont{Fertig}},
  \bibinfo{journal}{Phys. Rev. B}
  \textbf{\bibinfo{volume}{73}}, \bibinfo{eid}{235411}
  (\bibinfo{year}{2006}).

\bibitem[{\citenamefont{Roy et~al.}(1994)\citenamefont{Roy, Mendez, and
  Dominguez-Adame}}]{RoyMendez94}
\bibinfo{author}{\bibfnamefont{C.~L.} \bibnamefont{Roy}},
  \bibinfo{author}{\bibfnamefont{B.}~\bibnamefont{Mendez}}, \bibnamefont{and}
  \bibinfo{author}{\bibfnamefont{F.}~\bibnamefont{Dominguez-Adame}},
  \bibinfo{journal}{J. Phys. A} \textbf{\bibinfo{volume}{27}},
  \bibinfo{pages}{3539} (\bibinfo{year}{1994}).

\bibitem[{\citenamefont{Samsonov et~al.}(2003)\citenamefont{Samsonov,
  Pecheritsin, Pozdeeva, and Glasser}}]{SamsonovPecheritsin03}
\bibinfo{author}{\bibfnamefont{B.~F.} \bibnamefont{Samsonov}},
  \bibinfo{author}{\bibfnamefont{A.~A.} \bibnamefont{Pecheritsin}},
  \bibinfo{author}{\bibfnamefont{E.~O.} \bibnamefont{Pozdeeva}},
  \bibnamefont{and} \bibinfo{author}{\bibfnamefont{M.~L.}
  \bibnamefont{Glasser}}, \bibinfo{journal}{Eur. J. Phys.}
  \textbf{\bibinfo{volume}{24}}, \bibinfo{pages}{435}
  (\bibinfo{year}{2003}).

\bibitem[{\citenamefont{Calogeracos Dombey}(1999)}]{CalogeracosDombey99}
\bibinfo{author}{\bibfnamefont{A.}~\bibnamefont{Calogeracos}},
\bibinfo{author}{\bibfnamefont{N.}~\bibnamefont{Dombey}},
  \bibinfo{journal}{Contemp. Phys.} \textbf{\bibinfo{volume}{40}},
  \bibinfo{pages}{313} (\bibinfo{year}{1999}).

\bibitem[{\citenamefont{Datta}(1995)}]{Datta95}
\bibinfo{author}{\bibfnamefont{S.}~\bibnamefont{Datta}},
  \emph{\bibinfo{title}{Electronic transport in mesoscopic systems}}
  (\bibinfo{publisher}{Cambridge University Press},
  \bibinfo{address}{Cambridge}, \bibinfo{year}{1995}).

\bibitem[{\citenamefont{Katsnelson}(2007)}]{Katsnelson07}
\bibinfo{author}{\bibfnamefont{M.}~\bibnamefont{Katsnelson}},
  \bibinfo{journal}{Eur. Phys. J. B} \textbf{\bibinfo{volume}{57}},
  \bibinfo{pages}{225} (\bibinfo{year}{2007}).


\end{thebibliography}
\end{document}